\documentclass[12pt]{article}
\usepackage{epsfig}

\def\beq{\begin{equation}}
\def\eeq#1{\label{#1}\end{equation}}
\def\eeqn{\end{equation}}

\def\beqa{\begin{eqnarray}}
\def\eeqa#1{\label{#1}\end{eqnarray}}
\def\eeqan{\end{eqnarray}}

\let\bar=\overbar

\def\Dslash{\not{\hbox{\kern-4pt $D$}}}
\def\dslash{\not{\hbox{\kern-2pt $\del$}}}

\def\msb{{\bar{\ssstyle M \kern -1pt S}}}

\newcommand{\ls}{{\stackrel{\textstyle <}{_\sim}}}
\newcommand{\msun}{{M_{\odot}}}

\newcommand{\gcmt}{{\rm g/cm}^3}

\newcommand{\edrip}{\epsilon_{\rm drip}}
\newcommand{\ecrusti}{\epsilon_{\rm crust}}
\newcommand{\lsim}{\stackrel{\textstyle <}{_\sim}}

\newcommand{\pkgr}{P_{\rm K}}
\newcommand{\msec}{\rm msec}

\def\Title#1{\begin{center} {\Large {\bf #1} } \end{center}}

\begin{document}

\Title{From Neutron Stars to Strange Stars}

\bigskip\bigskip


\begin{raggedright}  

{\it {\underline {Fridolin Weber}}\index{Weber, F.}\\
Department of Physics\\
225 Nieuwland Science Hall\\
University of Notre Dame\\
Notre Dame, IN 46556-5670, USA} \\
\bigskip\bigskip
\end{raggedright}

\section{Introduction}

It is generally agreed that the tremendous densities reached in the
centers of neutron stars provide a high pressure environment in which
numerous particles processes are likely to compete with each
other. These processes range from the generation of hyperons to quark
deconfinement to the formation of kaon condensates and H-matter
\cite{weber99:book}.  Another striking possibility concerns the
formation of absolutely stable strange quark matter, a configuration
of matter even more stable than the most stable atomic nucleus,
iron. In the latter event all neutron stars would in fact be strange
(quark matter) stars \cite{madsen98:b}, objects largely composed of
pure strange quark matter, eventually enveloped in a thin nuclear
crust made up of ordinary, hadronic matter.

There has been much recent progress in our understanding of quark
matter, culminating in the discovery that if quark matter exists it
will be in a color superconducting state
\cite{alford98:a,rapp98:a,rajagopal01:a,alford01:a}. The phase diagram
of such matter appears to be very complex
\cite{rajagopal01:a,alford01:a}.  At asymptotic densities the ground
state of QCD with a vanishing strange quark mass is the color-flavor
locked (CFL) phase. This phase is electrically neutral in bulk for a
significant range of chemical potentials and strange quark masses
\cite{rajagopal01:b}. If the strange quark mass is heavy enough to be
ignored, then up and down quarks may pair in the two-flavor
superconducting (2SC) phase.  Other possible condensation patters are
the recently discovered CFL--$K^0$ phase \cite{bedaque01:a} and the
color-spin locked (2SC+s) phase \cite{schaefer00:a}.  The magnitude of
the gap energy lies between $\sim 50$ and $100$~MeV. Color
superconductivity thus modifies the equation of state (eos) at the
order $(\Delta / \mu)^2$ level, which is only a few percent. Such
small effects can be safely neglected in present determinations of
models for the eos of neutron stars and strange quark matter stars.
There has been much recent work on how color superconductivity in
neutron stars could affect their properties
\cite{rajagopal01:a,alford01:a,rajagopal00:a,alford00:a,alford00:b}.
These studies revealed that possible signatures include the cooling by
neutrino emission, the pattern of the arrival times of supernova
neutrinos, the evolution of neutron star magnetic fields, rotational
(r-mode) instabilities, and glitches in rotation frequencies.  In this
review I shall complement this list by reviewing several, most
intruigung astrophysical implications connected with the possible
absolute stability of strange quark matter.  (Surface properties of
strange matter are discussed in Usov's paper elsewhere in this
volume.)  This is followed by a discussion of two astrophysical
signals that may point at the existence of quark matter in both
isolated neutron stars as well as in neutron stars in low-mass x-ray
binaries (LMXBs). We recall that a convincing discovery of quark
matter in neutron stars would demonstrate that strange quark matter is
not absolutely stable, ruling out the absolute stability of strange
quark matter and the existence of strange quark stars, for it is not
possible for neutron stars to contain quark matter cores and strange
matte quark stars to both be stable \cite{rajagopal01:a}.

\goodbreak
\section{Nuclear crusts on strange matter stars}\label{sec:crust}

Since stars in their lowest energy state are electrically charge
neutral to very high precision, any net positive quark charge must be
balanced by leptons. As a general feature, there is only very little
need for leptons, since charge neutrality can be achieved essentially
among the quarks themselves. (This is specifically the case for
superconducting CFL quark matter in the asymptotic limit.)  If
electrons form a component of absolutely stable strange quark matter,
their presence is crucial for the possible existence of a nuclear
curst on such matter \cite{alcock86:a,kettner94:b}. The reason being
that the electrons, which are bound to strange matter by the Coulomb
force rather than the strong force, extend several hundred fermi
beyond the surface of strange matter.  Associated with this electron
displacement is a very strong electric dipole layer which can support,
out of contact with the surface of the strange matter, a crust of
nuclear material, which it polarizes.  The maximal possible density at
the base of the crust (inner crust density) is determined by neutron
drip, $\edrip=4.3\times 10^{11}~\gcmt$, at which neutrons begin to
drip out of the nuclei and form a free neutron gas.  Being
electrically charge neutral, the neutrons do not feel the repulsive
Coulomb force and hence would gravitate toward the quark matter core,
where they become converted into strange matter. Neutron drip thus
sets a strict upper limit on the crust's maximal inner density. The
actual value may be slightly smaller though \cite{althaus96:crust}.
The somewhat complicated situation of the structure of a strange
matter star with crust can be represented by a proper choice of eos
\cite{glen92:crust}, which consists of two parts. At densities below
neutron drip it is represented by the low-density eos of
charge-neutral nuclear matter, for which we use the
Baym-Pethick-Sutherland eos.  The star's strange matter core is
described by the bag model.

\goodbreak
\section{Complete sequences of strange matter stars}

\begin{figure}[htb] 
\begin{center}
\leavevmode
\epsfig{figure=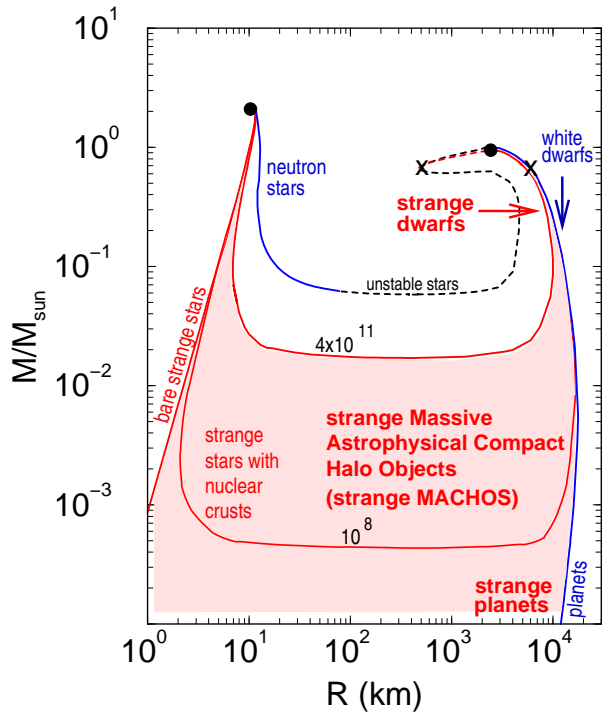,width=8.0cm}
\caption[]{Classification of stellar objects in the mass-radius
plane. Bare strange stars obey $M \propto R^3$. Nuclear crusts on
them give way to an expansive range of novel stellar
configurations (shaded area). These range from strange MACHOS to
strange dwarfs to strange planets. The labels $10^8$ and $4\times
10^{11}$ refer to inner crust densities in ${\rm g/cm}^3$.}
\label{fig:sequence}
\end{center}
\end{figure}

Since the nuclear crusts surrounding the cores of strange stars are
bound by the gravitational force rather than confinement, the
mass-radius relationship of strange matter stars with crusts is
qualitatively similar to the one of purely gravitationally bound
stars, neutron stars and white dwarfs, as illustrated in Fig.\
\ref{fig:sequence} \cite{weber93:b,glen94:a}.
The strange star sequences are computed for the maximal possible inner
crust density, $\ecrusti=\edrip$, as well as for an arbitrarily
chosen, smaller value of $\ecrusti=10^8~\gcmt$, which may serves to
demonstrate the influence of less dense crusts on the mass-radius
relationship \cite{weber93:b}.  From the maximum mass star (dot in the
upper left corner), the central density decreases monotonically
through the sequence in each case.  The stars located along the dashed
line represent unstable configurations \cite{weber93:b,glen94:a}.  The
fact that strange stars with crusts tend to possess somewhat smaller
radii than neutron stars implies smaller mass shedding (Kepler)
periods $\pkgr$ for the former. This is already indicated by the
classical expression $\pkgr=2\pi\sqrt{R^3/M}$ and carries over to the
full general relativistic determination of $\pkgr$
\cite{weber99:book},
\begin{equation} 
\pkgr = 2 \pi \, \Biggl( \omega + \frac{\omega^\prime}{2\psi^\prime}
+e^{\nu -\psi} \sqrt{ \frac{\nu^\prime}{\psi^\prime} +
\Bigl(\frac{\omega^\prime}{2 \psi^\prime}e^{\psi-\nu}\Bigr)^2 }
\; \Biggr)^{-1} \, .
\label{eq:okgr}
\end{equation}
This expression is to be computed simultaneously in combination with
Einstein's field equations for a rotating compact body
\cite{weber99:book},
\begin{equation} 
 {R}^{\kappa\lambda} \; - \; {1\over 2} \; g^{\kappa\lambda} \; {R} \;
 = \; 8\, \pi \; {T}^{\kappa\lambda}(\epsilon,P(\epsilon)) \, .
 \label{eq:einstein}
\end{equation}
It is found that, due to the smaller radii of strange stars, the
complete sequence of such objects, and not just those close to the
mass peak as is the case for neutron stars, can sustain extremely
rapid rotation \cite{weber93:b}.  In particular, model
calculations indicate for a strange star with a typical pulsar mass of
$\sim 1.45\,\msun$ Kepler periods in the range of $0.55 ~\msec \lsim 
\pkgr \lsim 0.8 ~ \msec$, depending on the thickness of the nuclear
curst and the bag constant \cite{glen92:crust,weber93:b}. This range
is to be compared with $\pkgr\sim 1~\msec$ obtained for neutron stars
of the same mass.

The minimum-mass configurations of the sample strange star sequences
in Fig.\ \ref{fig:sequence} have masses of about $\sim 0.017\, \msun$
(about 17 Jupiter masses) and $10^{-4}\,\msun$, depending on the
valuer of $\ecrusti$. For inner crust densities smaller than
$10^8~\gcmt$ one obtains stable strange matter stars that can be by
orders of magnitudes lighter than Jupiters. If abundant enough, these
light strange stars could be seen by the gravitational microlensing
searches \cite{alcock00:a}.  Strange stars located to the right of the
minimum mass configuration of each sequence consist of small strange
cores, typically smaller than about 3~km, surrounded by nuclear crusts
(ordinary white dwarf matter) that are thousands of kilometers
thick. Such objects are called strange dwarfs. Their cores have shrunk
to zero at the crossed points. What is left are ordinary white
dwarfs with central densities equal to the inner crust densities of
the former strange dwarfs.  A  stability analysis of strange
stars against radial oscillations \cite{weber93:b} shows that all
 strange dwarf sequences that terminate at stable ordinary white
dwarfs are stable against radial oscillations.  Strange stars that are
located to the left of the mass peak of ordinary white dwarfs (solid
dot in upper right corner), however, are unstable against
oscillations and thus cannot exist in nature.  So, in sharp
contrast to neutron stars and white dwarfs, the branches of strange
stars and strange dwarfs are stably connected with each other
\cite{weber93:b,glen94:a}. Finally we would like to stress that
strange dwarfs with $10^9~\gcmt<\ecrusti<4\times 10^{11}~\gcmt$ form
entire new classes of stars that contain nuclear material up to $\sim
4\times 10^4$ times denser than in ordinary white dwarfs of average
mass, $M\sim 0.6\,\msun$ (central density $\sim 10^7~\gcmt$).  The
entire family of such strange stars owes its stability to the strange
core.  Without the core they would be placed into the unstable region
between ordinary white dwarfs and neutron stars \cite{glen94:a}.

\begin{figure}[htb] 
\begin{center}
\leavevmode
\epsfig{figure=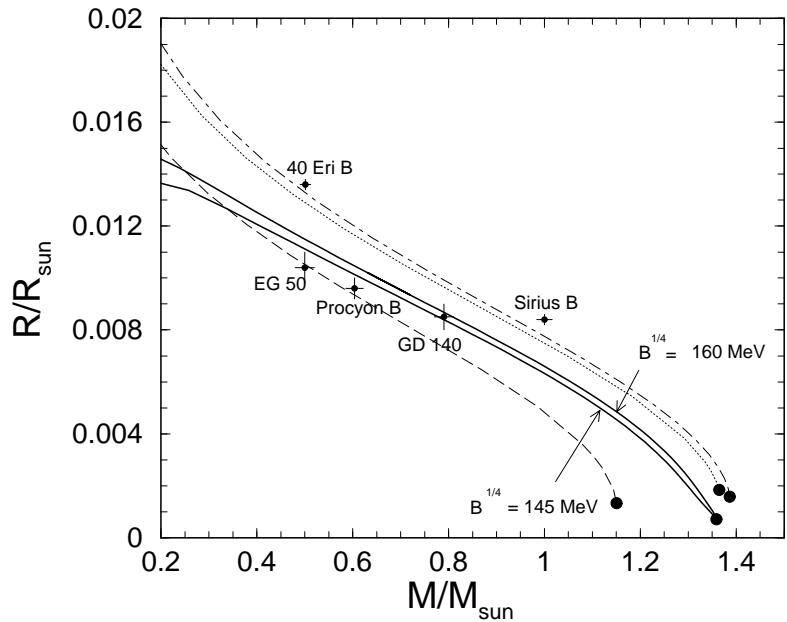, width=8.5cm,angle=-90}
\caption[]{Comparison of theoretical mass-radius relationships of
strange dwarfs ($\ecrusti=\edrip$) and several ordinary white dwarf stars
\cite{gorman01:a} ($^{12}{\rm C}$ dwarfs: dot-dashed, $^{24}{\rm Mg}$:
dotted, $^{56}{\rm Fe}$: dashed double-dotted) with data from the
Hipparcos project.}
\label{fig:mvsr}
\end{center}
\end{figure}

Until recently, only rather vague tests of the theoretical mass-radius
relation of white dwarfs have been possible. This has changed because
of the availability of new data emerging from the Hipparcos project
\cite{provencal98:a}. These data allow the first accurate measurements
of white dwarf distances and, as a result, establishing the mass-radius
relation of such objects empirically. Figure \ref{fig:mvsr} shows a
comparison of several data from the Hipparcos project with the mass-radius
relationships of strange dwarfs (solid lines) and ordinary white dwarfs
computed for different compositions.

\section{Post-glitch behavior of strange stars}\label{ssec:post}

Ordinary neutron stars older than a few months have crusts made of a
crystal lattice or an ordered inhomogeneous medium reaching from the
surface down to regions with a density of $2\times 10^{14}~ {{\rm
g/cm}^3}$.  This crust contains only a few percent of the total moment
of inertia. Strange stars, in contrast, can only support a crust with
a density below neutron drip ($4.3\times 10^{11}~ {{\rm g/cm}^3}$),
for reasons discussed in section~\ref{sec:crust}. Such a strange star
crust contains at most $\sim 10^{-5}$ of the total moment of
inertia. This is an upper bound, since the strange star may have no
crust at all, depending on its prior evolution. The difference in the
moment of inertia stored in the crust of neutron stars and strange
stars seems to pose significant difficulties for explaining the glitch
phenomenon observed in radio pulsars with models based on strange
stars. Glitches are observed as sudden speed-ups in the rotation rate
of pulsars. The fractional change in rotation rate $\Omega$ is
$\Delta\Omega/\Omega \approx 10^{-6} - 10^{-9}$, and the corresponding
fractional change in the spin-down rate $\dot\Omega$ is of order
$\Delta\dot\Omega/\dot\Omega \approx 10^{-2} - 10^{-3}$. In
\cite{glen92:crust} it has been demonstrated that strange stars with
crusts obey
\begin{equation}
  \frac{\Delta \Omega}{\Omega} \simeq \bigl( 10^{-5} ~ {\rm to} ~
  10^{-3} \bigr) ~ f \, , \qquad (0 < f < 1) \, ,
\label{eq:1.57.gli}
\end{equation} where  $f$  represents the
fraction of the crustal moment of inertia that is altered in the
quake, $|\Delta I| = f I_{\rm crust}$.  Since observed glitches have
relative frequency changes of $\Delta \Omega/ \Omega \simeq 10^{-9}$
to $10^{-6}$, a change in the crustal moment of inertia by less than
10\% would cause a giant glitch ($\Delta \Omega/ \Omega \sim 10^{-6}$)
even in the least favorable case.  Of course there remains the
question of whether there can be a sufficient build up of stress and
also of the recoupling of crust and core which involves the healing of
the pulsar period. This is probably a very complicated process that
does not simply involve the recoupling of two homogeneous substances.
For $\Delta\dot\Omega/\dot\Omega$ we established that 
\cite{glen92:crust}
\begin{equation}
  \frac{\Delta \dot{\Omega}}{\dot{\Omega}}  > (10^{-1} ~{\rm
    to}~ 10)~ f \, ,
\end{equation} yielding a small $f$ value as before, namely 
$f < 10^{-4} ~{\rm to}~ 10^{-1}$. 
 We have used measured values of the
ratio $(\Delta \Omega/\Omega) / (\Delta \dot{\Omega}/ \dot{\Omega})
\sim 10^{-6}~{\rm to}~10^{-4}$ for the Crab and Vela pulsars
respectively.  So the observed range of the
fractional change in $\dot{\Omega}$ is consistent with the
crust having the small moment of inertia calculated and the quake
involving only a small fraction, $f$, of that, just as in
Eq.\ (\ref{eq:1.57.gli}).  Nevertheless, without undertaking a
study of whether the nuclear solid crust on strange stars could
sustain a sufficient buildup of stress before cracking to account for
such a sudden change in relative moment of inertia, or whether the
healing-time and intervals between glitches can be understood, one
cannot say definitely that strange stars with a nuclear solid crust
can account for any complete set of glitch observations for a
particular pulsar. 

\section{Formation of strange dwarfs}

At present there is neither a well-studied model for the formation of
hypothetical strange dwarfs, nor exists a study that determines their
abundance in the universe. One possible scenario would be the
formation of strange dwarfs from main sequence progenitors that have
been contaminated with strange nuggets over their lifetimes. We recall
that the capture of strange matter nuggets by main sequence stars is
an inevitable consequence if strange matter were more stable than
hadronic matter \cite{madsen88:a} because then the Galaxy would be
filled with a flux of strange nuggets which would be acquired by every
object they come into contact with, i.e.\ planets, neutron stars,
white dwarfs, and main sequence stars. Naturally, due to the large
radii of the latter, they arise as ideal large-surface long
integration time detectors for the strange matter flux.  Nuggets that
are accreted onto neutron stars and white dwarfs, however, never reach
their centers, where the gravitational potential is largest, because
they are stopped in the lattice close to the surface due to the large
structural energy density there. This prevents such stars from
building up a cores of strange matter. The situation is different for
main sequence stars which are diffuse in comparison with neutron stars
and white dwarfs. In this case the accreted nuggets may gravitate to
the star's core, accumulate there and form a strange matter core that
grows with time until the star's demise as a main sequence star. 
Another plausible mechanism has to do with primordial strange matter
bodies.  Such bodies of masses between $10^{-2}$ and $1\,\msun$ may
have been formed in the early universe and survived to the present
epoch \cite{cottingham94:a}. Such objects will occasionally be
captured by a main sequence star and form a significant core in a
single and singular event.  The core's baryon number, however, cannot
be significantly larger than $\sim 5\!  \times\!
10^{31}~(M/\msun)^{-1.8}$ where $M$ is the star's mass.  Otherwise a
main sequence star is not capable of capturing the strange matter core
\cite{madsen98:b,madsen93:a}.  Finally we mention that in the very early
evolution of the universe lumps of hot strange matter will evaporate
nucleons which are plausibly gravitationally bound to the lump. The
evaporation will continue until the quark matter has cooled
sufficiently. Depending on the original baryon number of the quark
lump, a strange star or dwarf, both with nuclear crusts will have been
formed.

\section{Quark matter in isolated neutron stars}\label{sec:qmns}

Whether or not quark deconfinement occurs in neutron stars makes only
very little difference to their static properties, such as the range
of possible masses and radii, which renders the detection of quark
matter in such objects extremely complicated. This turns out to be
strikingly different for rotating neutron stars (i.e.\ pulsars) which
develop quark matter cores in the course of spin-down.  The reason
being that as such stars spin down, because of the emission of
magnetic dipole radiation and a wind of electron-positron pairs, they
become more and more compressed.  For some rotating neutron stars the
\begin{figure}[tb] 
\parbox[t]{7.0cm} 
{\leavevmode
\epsfig{figure=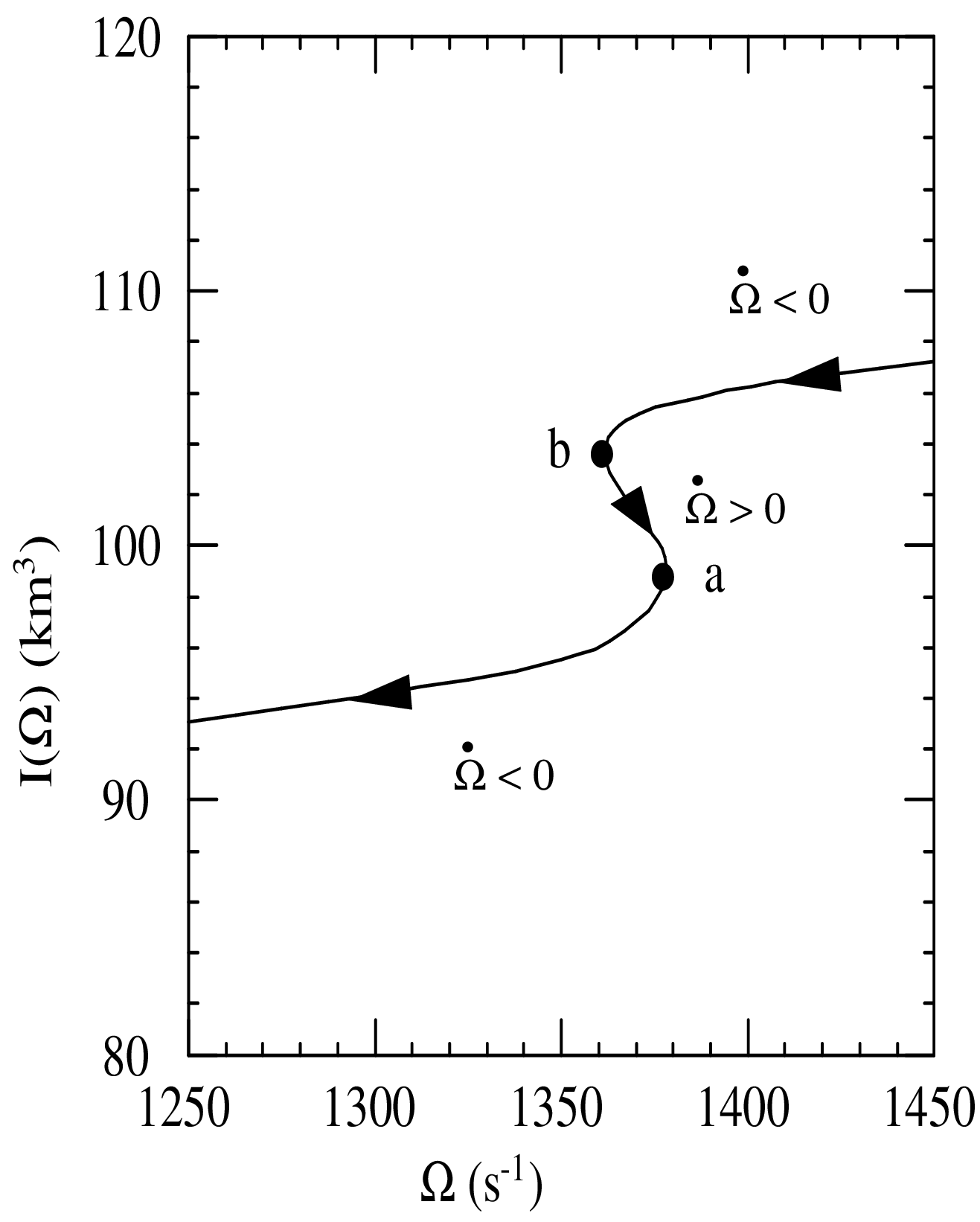,width=7.0cm,angle=-90}
{\caption[]{Moment of inertia versus frequency. The generation of
quark matter causes a ``backbending'' of $I$ for frequencies between $a$
and $b$ \protect{\cite{weber99:book,glen97:a}}.}
\label{fig:IOab}}}
\ \hskip 1.00cm \ 
\parbox[t]{7.0cm} 
{\leavevmode
\epsfig{figure=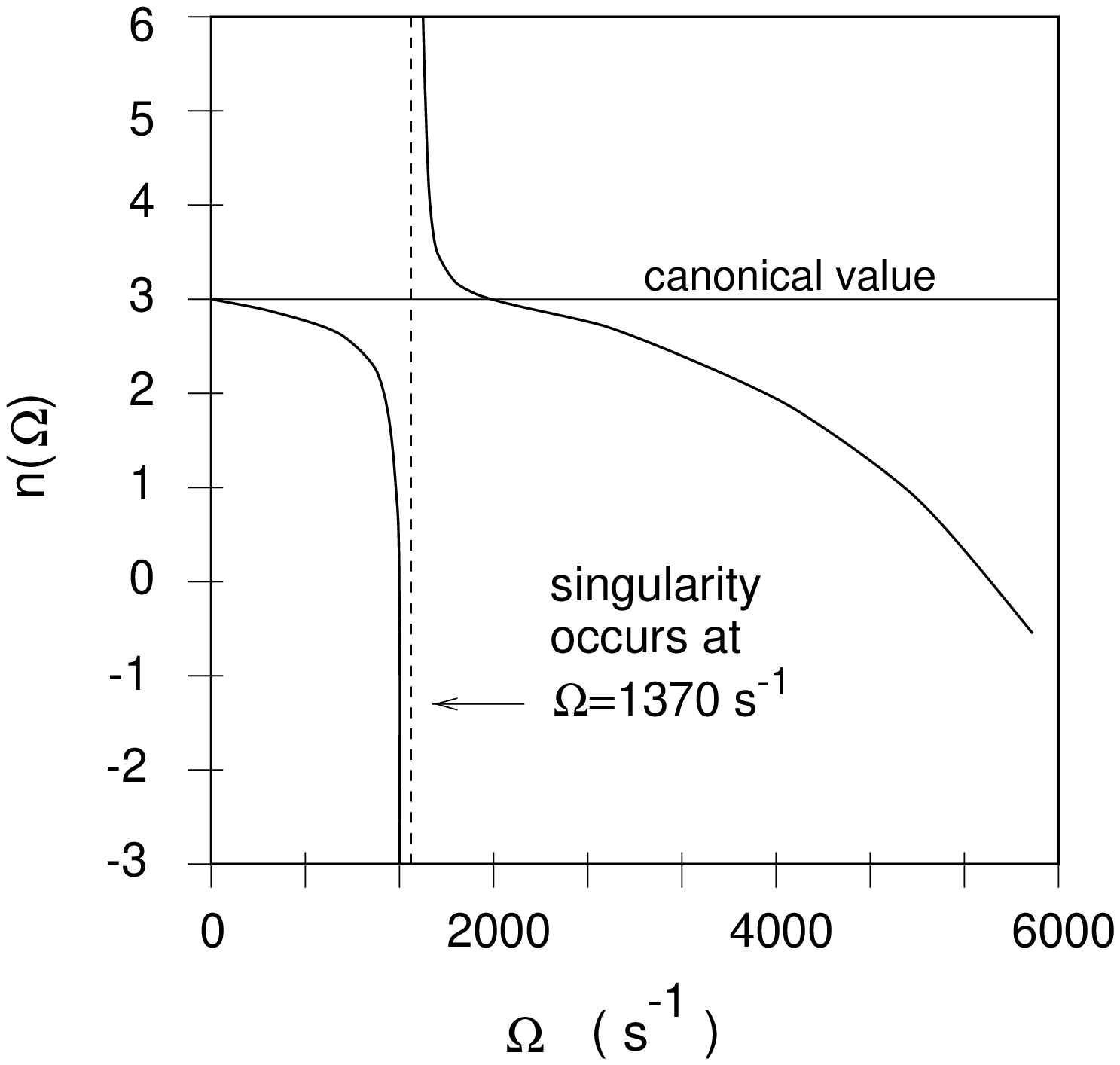,width=7.0cm,angle=-90}
\caption[]{Anomaly in braking index caused by generation of quark matter.}
\label{fig:nvso}}
\end{figure}
mass and initial rotational frequency may be just such that the
central density rises from below to above the critical density for
dissolution of baryons into their quark constituents. This effects
the star's moment of inertia dramatically \cite{glen97:a}, as shown in
figure~\ref{fig:IOab}. Depending on the ratio at which quark and
normal matter change with frequency, the moment of inertia can
decrease very anomalously, and could even introduce an era of stellar
spin-up lasting for $\sim 10^8$ years \cite{glen97:a}.  Since the
dipole age of millisecond pulsars is about $10^9$~years, we may
roughly estimate that about 10\% of the $\sim 25$ solitary millisecond
pulsars presently known could be in the quark transition epoch and
thus could be signaling the ongoing process of quark deconfinement!
Changes in the moment of inertia reflect themselves in the braking
index, $n$, of a rotating neutron star, as can be seen from $(I'\equiv
{\rm d}I/{\rm d}\Omega,~ I''\equiv {\rm d}^2I/{\rm d}\Omega^2)$
\begin{equation}  
  n(\Omega) \equiv \frac{\Omega\, \ddot{\Omega} }{\dot{\Omega}^2} = 3
    - \frac{ 3 \, I^\prime \, \Omega + I^{\prime \prime} \, \Omega^2 }
    {2\, I + I^\prime \, \Omega} \, .
\label{eq:index}
\end{equation}  The right-hand-side of this expression reduces to the 
well-known canonical constant $n=3$ if $I$ is independent of
frequency. Evidently, this is not the case for rapidly rotating
neutron stars, and it fails completely for stars that experience
pronounced internal changes (phase transitions) which alter the moment
of inertia significantly. Figure \ref{fig:nvso} illustrates this for
the moment of inertia of the neutron star of Fig.\ \ref{fig:IOab}.
Because of the changes in $I$, caused
\begin{figure}[htb] 
\parbox[t]{7.0cm} 
{\leavevmode 
\epsfig{figure=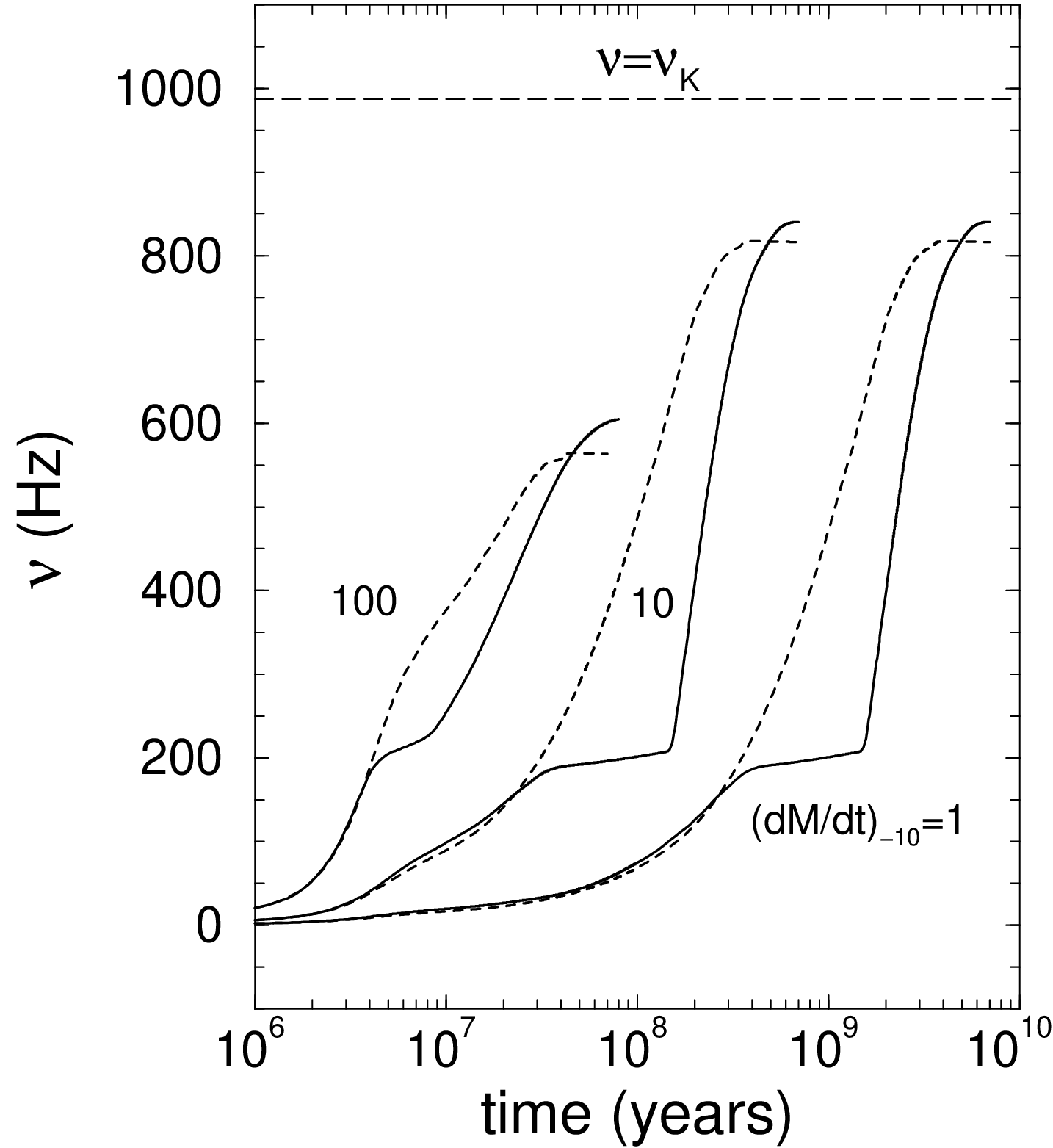,width=6.5cm}
{\caption[]{Evolution of spin frequencies of accreting x-ray neutron stars
with (solid curves) and without (dashed curves) quark deconfinement
\protect{\cite{glen01:a}}.  The spin plateau around 200~Hz signals the
ongoing process of quark re-confinement in the stellar centers
\cite{glen01:a}.}}}
\label{fig:nue_t}
\ \hskip 0.50cm \ 
\parbox[t]{7.0cm} 
{\leavevmode
\epsfig{figure=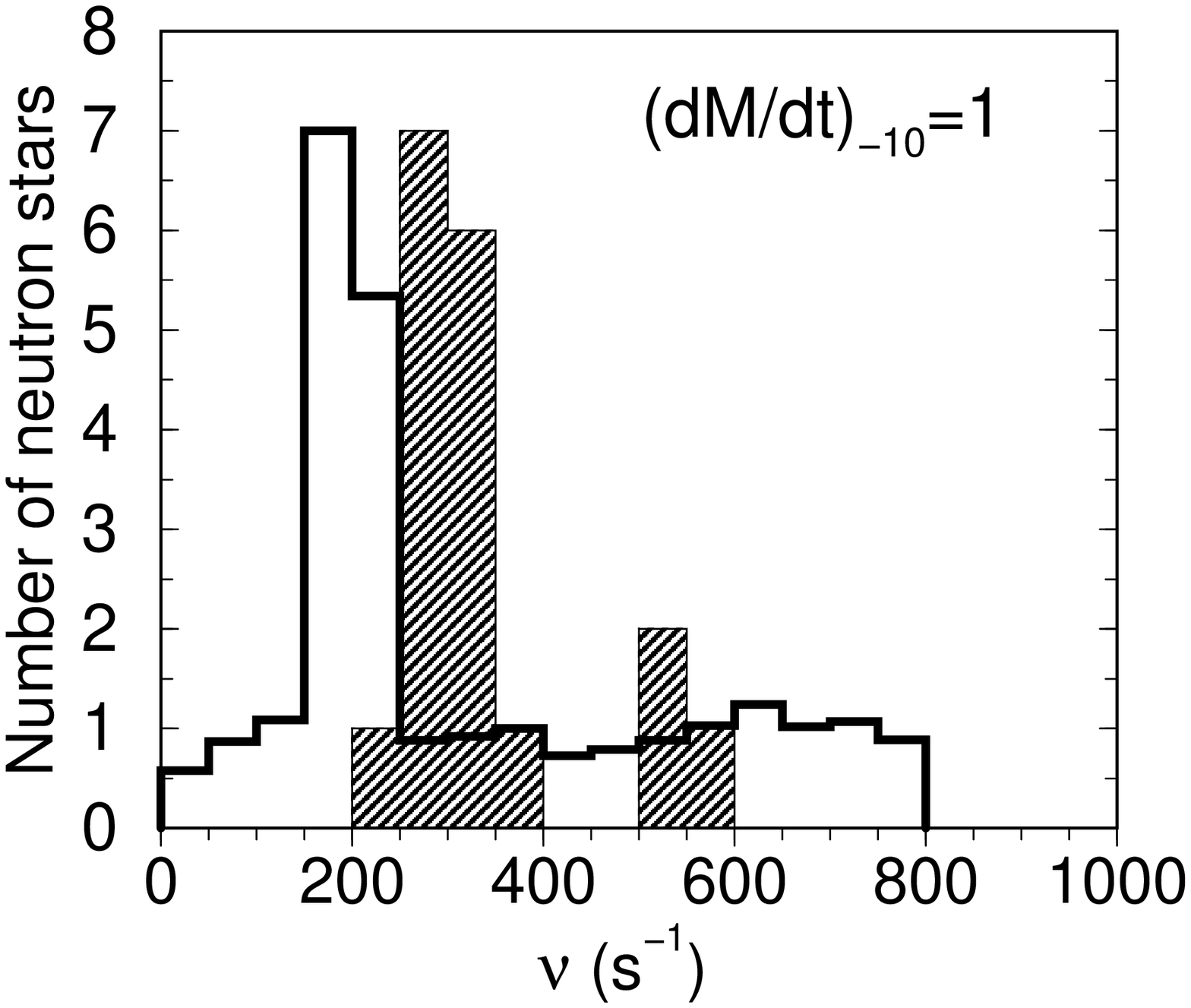,width=6.5cm}
\caption[]{Calculated spin distribution of x-ray neutron stars.  The
spike in the calculated distribution (unshaded diagram) corresponds to
the spinout of the quark matter phase. Otherwise the spike would be
absent. The shaded histogram displays the observed data
\cite{glen01:a}.}
\label{fig:histo}}
\end{figure}
by the gradual transition of hadronic matter into quark matter, the
braking index deviates dramatically from 3 at the transition
frequency, when pure quark matter is generated.  Such dramatic
anomalies in $n(\Omega)$ are not known for conventional neutron stars,
because their moments of inertia appear to vary smoothly with $\Omega$
\cite{weber99:book}. The future astrophysical observation of such
anomalies in the braking behavior of pulsars may thus be interpreted
as a signal for quark deconfinement in neutron stars.

\section{Quark matter in accreting x-ray neutron stars}
\label{sec:ano2}

Accreting x-ray neutron stars provide a very interesting contrast to
the spin-down of isolated neutron stars discussed in sect.\
\ref{sec:qmns}. These x-ray neutron stars are being spun up by the
accretion of matter from a lower-mass ($M \ls 0.4 \msun$), less-dense
companion.  If the critical deconfinement density falls within that of
canonical pulsars, quark matter will already exist in them but
will be ``spun out'' of x-ray stars as their frequency increases
during accretion.  This scenario has been modeled in
\cite{glen01:a,glen00:b} and will be discusses next.
The spin-up torque experienced by a neutron star causes a change in
the stars' angular momentum that is described by the relation
\begin{eqnarray}
{{dJ} \over {d t}} = {\dot M} {\tilde l}(r_{\rm m}) - N(r_{\rm c}) \,
,
\label{eq:dJdt}
\end{eqnarray}
where $\dot{M}$ denotes the accretion rate and $ {\tilde l}(r_{\rm m})
= \sqrt{M r_{\rm m}}$ is the angular momentum added to the star per
unit mass of accreted matter. The quantity $N(r_{\rm c}) = \kappa
\mu^2 r_{\rm c}^{-3}$ stands for the magnetic plus viscous torque
terms, with $\mu \equiv R^3 B$ the star's magnetic moment.  The
quantities $r_{\rm m}$ and $r_{\rm c}$ denote the radius of the inner
edge of the accretion disk and the co-rotating radius, respectively,
and are given by $(\xi \sim 1)$ $r_{\rm m} = \xi r_{\rm A}$ and
$r_{\rm c} = \left( M \Omega^{-2} \right)^{1/3}$.  Accretion will be
inhibited by a centrifugal barrier if the neutron star's magnetosphere
rotates faster than the Kepler frequency at the magnetosphere. Hence
$r_{\rm m} < r_{\rm c}$, otherwise accretion onto the star will cease.
The Alf\'en radius $r_{\rm A}$, where the magnetic energy density
equals the total kinetic energy of the accreting matter is defined by
$r_{\rm A} = {\mu^{4/7}} ({2 M \dot{M}^2})^{-1/7}$.  The rate of
change of a star's angular frequency $\Omega = 2 \pi / \nu$ $(=
J/I)$ then follows from Eq.\ (\ref{eq:dJdt}) as
\begin{equation}
  I(t) {{d\nu(t)} \over {d t}} = { {{\dot M} {\tilde l}(t)}\over{2 \pi}
  } - \nu(t) {{dI(t)}\over{dt}} - \kappa \, \mu(t)^2 \, r_{\rm
  c}(t)^{-3} \, ,
\label{eq:dOdt.1}
\end{equation} with the explicit time dependences as indicated.  Evidently, the
second term on the right-hand-side of Eq.\ (\ref{eq:dOdt.1}) depends
linearly on $\Omega$ while the third terms grows quadratically with
$\Omega$. The temporal change of the moment of inertia of accreting
neutron stars which undergo phase transitions is crucial
\cite{glen01:a,glen00:b}. This also renders the calculation of the
moment of inertia, given by \cite{weber99:book}
\begin{eqnarray}
  I(t) = 2\pi \int_0^\pi d\theta \int_0^{R(\theta;t)} dr \,
  e^{\lambda+\mu+\nu+\psi} {{\epsilon + P}\over{e^{2\nu - 2\psi} -
  (\omega - \Omega)^2}} {{\Omega - \omega} \over {\Omega}} \, ,
\label{eq:I} 
\end{eqnarray} 
very cumbersome, for each quantity on the right-hand-side varies
accordingly during stellar spin-up. (We assume rigid body rotation,
which renders the star's frequency $\Omega$ constant throughout the
star.) The solution of equation (\ref{eq:dOdt.1}) is shown in Fig.\
\ref{fig:nue_t}. The result is most striking. One sees that quark
matter remains relatively dormant in the stellar core until the star
has been spun up to frequencies at which the central density is about
to drop below the threshold density at which quark matter exists. As
known from Fig.\ \ref{fig:IOab}, this manifests itself in a
significant increase of the star's moment of inertia. The angular
momentum added to a neutron star during this phase of evolution is
therefore consumed by the star's expansion, inhibiting a further
spin-up until the quark matter has been converted into a mixed phase
of matter made up of hadrons and quarks.  Such accreters, therefore,
tend to spend a greater length of time in the critical frequencies
than otherwise. There will be an anomalous number of accreters that
appear at or near the same frequency, as shown in Fig.\
\ref{fig:histo}. This is what was found recently with the Rossi x-ray
Timing Explorer (shaded area in Fig.\ \ref{fig:histo}).  Quark
deconfinement constitutes an attractive explanation for this anomaly
\cite{glen01:a,glen00:b,poghosyan01:a}, though alternative
explanations were suggested too \cite{bildsten98:a,andersson00:a}.

\medskip {\bf Acknowledgments:} I am grateful to the organizers of
 Compact Stars in the QCD Phase Diagram, R.\ Ouyed and F.\ Sannino.

\def\Discussion{
\setlength{\parskip}{0.3cm}\setlength{\parindent}{0.0cm}
     \bigskip\bigskip      {\Large {\bf Discussion}} \bigskip}
\def\speaker#1{{\bf #1:}\ }
\def\endDiscussion{}

\vspace{-0.5cm}

\Discussion

\speaker{S. Balberg (Hebrew University)} The ``anomalous'' behavior in spin
down (up) of an isolated (pair in a LMXB) neutron star attributed to the quark
phase is really a feature of the equation of state, not of quark matter in
itself. In other words, we need a relatively sharp feature in the equation of
state to get a significant change in the moment of inertia, which would, in
principle be due to other phenomena as well.

\speaker{Weber} One needs a pronounced feature in the equation of
state in order to get an anomaly in the breaking index
(spin-up). Hyperons and meson condensates lead to features in the
equation of state that seem to be way to week to cause such a striking
anomaly \cite{weber99:book}. A softening alone, as caused by hyperons
and/or bosons, is not sufficient.  What is really required is a
softening followed by a gradual stiffening at higher densities which
is naturally obtained for hadron-quark matter gradually compressed to
higher densities.

\speaker{H. Heiselberg (NORDITA)} How does $M(R)$ look for your quark matter
star? 

\speaker{Weber} Glendenning and Kettner demonstrated the existence of
non-identical neutron star twins, depending on whether or not the
low-density sequence of neutron stars terminates at central densities
that fall close to the end of the mixed phase of quarks and hadrons
\cite{glen00:e}.

\endDiscussion
 
\end{document}